\documentclass[12pt]{article}

\begin{document}
\title{{\bf Massless particle in 2d spacetime with 
constant curvature}\thanks{Corresponding author: W\l odzimierz 
Piechocki, Theory Division (Zd.P8), 
So\l tan Institute for Nuclear Studies, Ho\.{z}a 69, 00-681 Warsaw, 
Poland, E-mail: piech@fuw.edu.pl,  Fax: (48 22) 621 60 85}}
\author{George Jorjadze$^\dag$ and 
W{\l}odzimierz Piechocki$^\ddag$ \\
{\small $^\dag$Razmadze Mathematical Institute, 
380093 Tbilisi, Georgia }\\
{\small $^\ddag$So{\l}tan Institute for Nuclear Studies,  
00-681 Warsaw, Poland }}
\maketitle

\begin{abstract}
\noindent
We consider dynamics of  massless particle in 2d 
spacetimes with constant curvature. We analyze different examples 
of spacetime. Dynamical integrals are constructed from spacetime 
symmetry related to $sl(2.{\bf R})$ algebra. Mass-shell condition 
restricts dynamical integrals to a cone (without vertex) which 
defines physical-phase space. We parametrize the cone by canonical 
coordinates. Canonical quantization with definite choice of operator 
ordering leads to unitary irreducible representations of 
$SO_\uparrow (2.1)$ group.

\noindent  PACS: 0460M, 0220S, 0240K

\end{abstract}

\pagebreak
\section{Introduction}

We analyze classical and quantum dynamics of a massless particle 
in two dimensional spacetime with constant curvature $R_0\neq 0.$ 
We consider the cases when spacetime is: (i) hyperboloid with 
$R_0<0,$ (ii) half-plane with $R_0<0$ and (iii) stripe with 
$R_0 > 0.$ Presented results complete our analysis of 
dynamics of a particle with non-zero mass $m_0$ in 2d curved 
spacetime 
[1,2]. Taking formally $m_0\rightarrow 0$ leads to some
singularities both at classical and quantum levels. Thus, 
the massless case needs separate treatment. As the result  
we get  clear picture of the role played 
by topology and global symmetries of spacetime in the procedure 
of canonical quantization.

Dynamics of a massless particle in 
gravitational field $g_{\mu\nu}(x^0,x^1)$ is defined by the action [3]
\begin{equation}
S=\int L(\tau )~d\tau,~~~~
L(\tau):=-\frac{1}{2\lambda (\tau )}~g_{\mu\nu}(x^0(\tau),
x^1(\tau))\;\dot{x}^\mu(\tau)\dot{x}^\nu(\tau),
\end{equation}
where $\tau$ is an evolution parameter, $\lambda $
plays the role of Lagrange multiplier and $\dot{x}^\mu := dx^\mu/d\tau$. 
It is assumed that $\lambda > 0$ and $\dot x^0>0$.

The action (1.1) is invariant under reparametrization $\tau
\rightarrow f(\tau), \lambda (\tau ) 
\rightarrow \lambda (\tau )/\dot f(\tau )$. 
This gauge symmetry leads to the constrained dynamics in the 
Hamiltonian formulation [4]. The constraint reads
\begin{equation}
\Phi :=g^{\mu\nu}p_\mu p_\nu =0,
\end{equation} 
where $p_\mu := \partial L/\partial\dot{x}^\mu$ are canonical 
momenta (we use units with $c=1=\hbar$). 

As in [1,2] we use the gauge invariant description.
In the case of massive particle the set of spacetime trajectories 
can be considered as a physical phase-space of the system.
This set has  natural symplectic structure, which can be used
for  quantization. The spacetime trajectories of a massless
particle has no symplectic structure (this set is only one
dimensional, since particle velocity is fixed). For the gauge invariant
description we use the dynamical integrals constructed from
the global symmetries of spacetime. 

\setcounter{equation}{0}
\section{Dynamics on hyperboloid}  

Let ($y^0,y^1,y^2$) be the standard coordinates on 3d Minkowski 
space with the metric tensor $\eta_{ab}=diag(+,-,-)$.
A one-sheet hyperboloid $\mathbf{H}$ is defined by
\begin{equation}
-(y^0)^2+(y^1)^2+(y^2)^2=m^{-2},
\end{equation}
where $m>0$ is a fixed parameter.
$\mathbf{H}$ has a constant curvature $R=-2m^2~$ (see [5]).

Any 2-dimensional Lorentzian 
manifold with constant curvature $R_0$ can be described (locally)
by the conformal metric tensor [5]
\begin{equation}
g_{\mu\nu}(X)= \exp\varphi(X)\;\left( \begin{array}{cr}
  1&0\\0&-1 \end{array}\right),~~~~~~~X:=(x^0,x^1),
\end{equation}
where the field $\varphi(X)$ satisfies the Liouville equation [6]
\begin{equation}
(\partial^2_0 -\partial^2_1)\varphi (X) + R_0\exp\varphi (X)=0 .
\end{equation}

Making use of the parametrization
\begin{eqnarray}
y^0=-\frac{\cot m\rho}{m},~~~~y^1=\frac{\cos m\theta}{m\sin 
m\rho},~~~~y^2=\frac{\sin m\theta}{m\sin m\rho},\nonumber \\
{\mbox {where}}~~~~ \rho \in ]0,\pi/m[,~~~\theta \in [0,2\pi/m[ 
~~~~~~~~~~~~~~~~~~~~~
\end{eqnarray}
we get the conformal form (2.2), with
\begin{equation}
\varphi =-\log \sin^2 m\rho,
\end{equation}
where the spacetime coordinates $x^0$ and $x^1$ are identified with 
the parameters $\rho$ and $\theta$, respectively.
The function (2.5) 
satisfies the Liouville equation (2.3) for $R_0 =-2m^2$.

The Lagrangian (1.1) in this case reads
\begin{equation}
L= -  \frac{\dot{\rho}^2-\dot{\theta}^2}
{2\lambda\sin^2 m\rho}~.
\end{equation}

The hyperboloid (2.1) is invariant under 
the Lorentz transformations, i.e., $SO_\uparrow (2,1)$ is the 
symmetry group of our system. The corresponding infinitesimal 
transformations (rotation and two boosts) are
\begin{eqnarray}
(\rho,\theta)\longrightarrow(\rho,\theta+\alpha_0/m),~~~~~~~~~~~~~~~~~~~~
\nonumber \\
(\rho,\theta)\longrightarrow(\rho-\alpha_1/m~ \sin m\rho\sin 
m\theta,\theta+\alpha_1/m~\cos m\rho\cos m\theta),\nonumber \\
(\rho,\theta)\longrightarrow(\rho+\alpha_2/m~\sin m\rho\cos m
\theta,\theta+\alpha_2/m~\cos m\rho\sin m\theta).
\end{eqnarray}
The dynamical integrals for (2.7) read
\begin{eqnarray}
J_0=\frac{p_{\theta}}{m},~~~~~J_1=-\frac{p_\rho}{m}\sin 
m\rho\sin 
m\theta +\frac{p_\theta}{m}\cos m\rho\cos m\theta, \nonumber \\
J_2=\frac{p_\rho}{m}\sin m\rho\cos m\theta 
+\frac{p_\theta}{m}\cos m\rho\sin m\theta,~~~~~~~~~~~~
\end{eqnarray}
where $p_\theta :=\partial L/\partial\dot{\theta},~p_\rho :=
\partial L/\partial\dot{\rho}$ are canonical momenta.

Since $J_0$ is connected with space translations (see (2.7)),
it defines particle momentum $p_{\theta}=mJ_0$.

It is clear that the dynamical integrals (2.8) satisfy the commutation
relations of $sl(2.\bf {R})$ algebra
\begin{equation}
\{ J_a , J_b \} =\varepsilon_{abc}\eta^{cd}J_d ,
\end{equation}
where $\eta^{cd}$ is the Minkowski metric tensor and 
$\varepsilon_{abc}$
is the anti-symmetric tensor with $\varepsilon_{012}=1$.

The mass shell condition (1.2) takes the form $p_\rho^2 -p_\theta^2 = 0$
and it leads to the relation
\begin{equation}
J_0^2 - J_1^2 - J_2^2 = 0 .
\end{equation}
Eq. (2.10) defines two cones. The singular point of the cones
$J_0=0=J_1=J_2$ should be removed, since it corresponds to the
massless particle with zero momentum ($p_\rho =0=p_\theta $),
which does not exist.

Thus, the dynamical integrals (2.8) define the physical
phase-space of the system and it consists of
two disconnected cones $\cal C_+$ and $\cal C_-$, for
$J_0 >0$ and $J_0<0$, respectively.

According to (2.8) and due to $p_\rho < 0$ 
(since $\dot\rho >0$ and $\lambda >0$) the trajectories satisfy
the equations
\begin{equation}
J_a y^a =0,~~~~~~~~
J_1y_2 - J_2y_1 = \frac{p_\rho}{m^2} = -\frac{|J_0|}{m}  .
\end{equation}
Each point ($J_a$) of the cone
$\cal C_+$  or $\cal C_-$ defines uniquely  the trajectory 
on the hyperboloid $\mathbf{H}$. The trajectories (2.11) are 
straight lines in 3d Minkowski space. Hence, the set of trajectories 
is the set of generatrices of the hyperboloid (2.1). For $J_0>0$ we 
get the `right' moving particle with $\dot\theta >0$, while for 
$J_0<0$ we have the `left' moving one, with $\dot\theta <0$.
Both cones, ${\cal C}_+$ and ${\cal C}_-$, are invariant under
$SO_\uparrow (2.1)$ transformations. 

To quantize the system, we consider the cones ${\cal C}_+$ and 
${\cal C}_-$ separately.
We parametrize ${\cal C}_+$ as follows
\begin{equation}
J_0 = \frac{1}{2}(p^2 + q^2),~~~~J_1 = \frac{p}{2}~
\sqrt{p^2 + q^2},~~~~~J_2 = \frac{q}{2}~\sqrt{p^2 + q^2},
\end{equation}
where $(0,0)\neq (p,q)\in {\bf {R}}^2 $.

\noindent 
It is easy to see that (2.12) defines the one-to-one map
from the plane without the origin to
${\cal C}_+$.
The canonical commutation relation
$\{p,q\}=1$ provides (2.9).

For quantization of  ${\cal C}_+$ system
we introduce the creation-annihilation operators 
$a^\pm :=(\hat p \pm i\hat q)/\sqrt 2$
and choose the definite operator ordering in (2.12). This ordering 
is defined by the following requirements:

a) the operators
$\hat J_a$
are self-adjoint,

b) they generate global 
$SO_\uparrow (2.1)$
transformations,

c) the spectrum of 
$\hat J_0$
is positive,

d) the Casimir number operator 
$\hat C:=\hat J_0^2-\hat J_1^2 -\hat J_2^2$
is zero.

This leads to the following expressions
\begin{equation}
\hat J_0 = a^-a^+,~~~~\hat J_+ = a^+\sqrt{a^-a^++1},
~~~~\hat J_- = \sqrt{a^-a^++1}~a^-,
\end{equation}
where $\hat J_\pm=\hat J_1\pm\hat J_2$. The 
creation and annihilation
operators are
defined in the Fock space in the standard way
$$
a^+|n\rangle =\sqrt{n+1}~|n+1\rangle ,~~~~~~~~~
a^-|n\rangle =\sqrt{n}~|n-1\rangle ,
$$
where the vectors $|n\rangle$, $(n\geq 0)$ form
the basis of the corresponding Hilbert space
${\cal H}_+$.

The states $|n\rangle$ are the eigenstates of
$\hat J_0$
\begin{equation}
\hat J_0 |n\rangle=(n+1) |n\rangle ,
\end{equation}
and from (2.13) we  get
\begin{equation}
\hat J_+|n\rangle = \sqrt{(n+1)(n+2)}~|n+1\rangle,
~~~~\hat J_-|n\rangle = \sqrt{n(n+1)}~|n-1\rangle .
\end{equation}
Eqs.(2.13)-(2.15) define the unitary irreducible representation (UIR) 
of the group $SO_\uparrow (2.1)$.
We identify this representation with the representation of 
$D_1^+$ of the discrete series of 
$SL(2.{\bf R})$ group [7].

Quantization of the case 
${\cal C}_-$
can be done in the same way. The corresponding Hilbert space
${\cal H}_-$
is again the Fock space, but we have to make the following
replacements:
$\hat J_\pm \rightarrow \hat J_\mp$ and $\hat J_0\rightarrow -\hat J_0$.
As a result we get the representation $D_1^-$.

The Hilbert space of the whole system is
${\cal H}={\cal H}_+\oplus{\cal H}_-$ and the corresponding representation
$D_1^+\oplus D_1^-$
describes the $SO_\uparrow (2.1)$ symmetry of the quantum system.

It is interesting to mention the following:

At the classical level our system can be obtained from the massive
case in the limit 
$m_0\rightarrow 0$
(where $m_0$ is a particle mass)
and by removing the singular point of the physical phase-space.
\noindent
The quantum theory of the massive case was considered recently in [1].
The corresponding representation is defined by the operators
\begin{eqnarray}
\hat J_0 \psi_n = n\psi_n,~~~~\hat J_+ \psi_n =
\sqrt{n^2+n+a^2}~\psi_{n+1},
\nonumber \\
\hat J_- \psi_n =\sqrt{n^2 - n + a^2}~\psi_{n-1},~~~~~~~~~~~~~~~~
\end{eqnarray}
where $\psi_n :=\exp~in\phi$ ($n\in Z$) form the basis 
of the Hilbert space
$L_2(S^1)$ and $a=m_0/m$. 
\noindent
For $a=0$ ($m_0=0$) this representation terns into 
$D_1^+\oplus A\oplus D_1^-$, where $A$ is a one dimensional
trivial representation on the vector $\psi_0$.
By removing $A$ we get the quantum theory of the massless case.

The momentum of a quantum particle 
$\hat p_{\theta}=m\hat J_0$ 
can take only discrete values
$P_n=mn$, where $n$ is a nonzero integer. 

\setcounter{equation}{0}
\section{Dynamics on half-plane}

Let us consider the Liouville field 
$\varphi = -2\log m|x^0|$,
given on a plane $(x^0, x^1)$. This field
defines constant spacetime curvature $R_0=-2m^2$ (see (2.3)).
The Lagrangian (1.1) in this case reads
\begin{equation}
L = - {\frac{2\dot x^+ \dot x^-}{\lambda m^2(x^+ + x^-)^2}},
\end{equation}
where $x^\pm :=x^0 \pm x^1$. 
It is assumed that
$\dot x^0 > 0$, which leads to $p_++p_- < 0$.

Formally, (3.1) is invariant under the fractional-linear
transformations
\begin{equation}
x^+ \rightarrow \frac{ax^+ + b}{cx^+ + d}, ~~~~~
x^- \rightarrow \frac{ax^- - b}{-cx^- + d},~~~~~~ ad-bc =1.
\end{equation}
Thus, formally, $SL(2.\bf {R})/Z_2$ (which is isomorphic
to $SO_\uparrow (2.1)$) is the symmetry of our system.
The transformations (3.2) are well defined on the plane
only for $c=0$. The corresponding transformations with $c=0$
form the group of dilatations and translations (along $x^1$),
which is a global symmetry of the considered spacetime.

The infinitesimal transformations for (3.2) are
\begin{equation}
x^\pm \rightarrow x^\pm \pm \alpha_0,~~~~~~ 
x^\pm \rightarrow x^\pm + \alpha_1 x^\pm,~~~~~~
x^\pm \rightarrow x^\pm \pm \alpha_2 (x^\pm)^2
\end{equation}
and the corresponding dynamical integrals read
\begin{equation}
P = p_+ - p_-,~~~~~K = p_+x^+ + p_-x^-,~~~~M = p_+(x^+)^2
- p_-(x^-)^2 ,
\end{equation}
where $p_\pm =\partial L/\partial \dot x^\pm$.

The dynamical integrals (3.4)
satisfy again the commutation relations (2.9) with
\begin{equation}
J_0 =\frac{1}{2}(P + M), ~~~~~~
J_1 =\frac{1}{2}(P - M), ~~~~~~ J_2 = K .
\end{equation}
The mass-shell condition (1.2) 
leads to $p_+=0$ for $P>0$ and $p_-=0$ for $P<0$.
Due to (3.4) and the mass-shell condition, we have 
\begin{equation}
K^2 -PM =0
\end{equation}
 and the trajectories read
\begin{equation}
x^1 - \epsilon (P)x^0 =\frac{K}{P},~~~~~\mbox
{where}~~~~~ \epsilon (P) =\frac{P}{|P|}.
\end{equation} 
 
It is clear that $x^0 =0$ is the singularity line 
in the spacetime. 
In the massive case this singularity leads to the dynamical
ambiguities [2,8]. However, the dynamics of the massless particle 
is defined uniquely due to (3.7).

The physical phase-space is defined by two cones (3.6)
without the line corresponding to $P=0$. Thus, we have
two disconnected 
parts: ${\cal P}_+$ with $P> 0$ and ${\cal P}_-$ with $P<0$.
Both cones (${\cal P}_+$ and ${\cal P}_-$)
are invariant under 
dilatations and translations generated by the dynamical
integrals $K$ and $P$. 
But, the transformations generated by $M$
are not defined globally.
Thus, the physical phase-space has the same symmetry as the spacetime.

Let us quantize the system corresponding to ${\cal P}_+$ case 
(the case ${\cal P}_-$ can be done in the same way).

We parametrize ${\cal P}_+$ as follows [9]
\begin{equation}
P=p,~~~~~~K = pq ,~~~~~~~~M = pq^2  ,
\end{equation}
 where ($p,q$) are the coordinates on half-plane with $p >0$.
The canonical commutation relation $\{p,q\}=1$ provides
the commutation relations of $sl(2.\bf {R})$ algebra 
\begin{equation}
\{P, K\} =P, ~~~~~\{P, M\} =2K, ~~~~~\{K, M\} =M~. 
\end{equation} 
In the `p-representation' ( $\hat q =i\partial_p$ )  we choose
the operator ordering for $\hat K$ and $\hat M$
by the following requirements:

a) $\hat K$ and $\hat M$ are self-adjoint,

b) $\hat P$, $\hat K$ and $\hat M$ satisfy the commutation relations
corresponding to (3.9),

c) the Casimir number operator $\hat C =
1/2(\hat P\hat M +\hat M\hat P)-\hat K^2$ equals zero.

As a result we get
\begin{equation}
\hat P = p ,~~~~~
\hat K = i (p\partial_p +\frac{1}{2}),~~~~~
\hat M= - p\partial^2_p - \partial_p +\frac{1}{4p}.
\end{equation} 
The operators defined by (3.10) have continuous spectrum. 

\noindent
The operator $\hat J_0=(\hat P+\hat M)/2$ has the discrete spectrum
\begin{equation}
\hat J_0 \psi_n (p) =\lambda_n \psi_n (p),
\end{equation}
 \begin{equation}
 \psi_n (p) = \sqrt p~\exp {(-p)} L^1_n(2p),~~~~
 \lambda_n =n+1,~~~~ n\geq 0,
\end{equation} 
where $L^1_n(x)$ are the Laguerre polynomials defined by [10]
\begin{equation}
L^1_n(x) = \sum_{k=0}^{n-1} (-)^k\frac{n!(n+1)\cdots
(k+2)}{k!(n-k)!}x^k
+(-)^nx^n .
\end{equation}
Eqs. (3.11)-(3.13) show that our representation is unitarily equivalent
to $D_1^+$ representation (see (2.14) and (2.15)). 
The scheme presented here can be generalized to include 
other representations of the discrete series of $SL(2.{\bf R})$ 
group [9].

\setcounter{equation}{0}
\section{Dynamics on stripe}

Let us consider the spacetime to be a stripe
\begin{equation}
{\mathcal {S}}:=\{(t,x)~|~t\in {\bf {R}}, ~x\in] 0 ,\pi /m [ \},
\end{equation} 
with the conformal metric tensor (2.2) and the Liouville field
\begin{equation}
\varphi (t, x)=-\log{\sin^2 mx}.
\end{equation}
It defines the spacetime with constant positive curvature
$R=2m^2$.

The Lagrangian (1.1) in this case reads
\begin{equation}
L= - \frac{\dot{t}^2-\dot{x}^2}
{2\lambda\sin^2 mx}
\end{equation}
and it is invariant under the action of the  
universal covering
group $\widetilde {SL}(2.{\bf R})$.
The corresponding infinitesimal 
transformations 
\begin{eqnarray}
(t,x)\longrightarrow(t-\alpha_0/m, x),~~~~~~~~~~~~~~~~~~~~
\nonumber \\
(t,x)\longrightarrow(t-\alpha_1/m~ \cos mx\cos mt,
x+\alpha_1/m~\sin mx\sin mt),\nonumber \\
(t,x)\longrightarrow(t-\alpha_2/m~\cos mx\sin mt,
x-\alpha_2/m~\sin mx\cos mt)~
\end{eqnarray}
lead to the dynamical integrals 
\begin{eqnarray}
J_0=-\frac{p_{t}}{m},~~~~~J_1=-\frac{p_t}{m}\cos 
mx\cos mt +\frac{p_x}{m}\sin mx\sin mt, \nonumber \\
J_2=-\frac{p_t}{m}\cos mx\sin mt 
-\frac{p_x}{m}\sin mx\cos mt,~~~~~~~~~~~~
\end{eqnarray}
which satisfy the commutation relations (2.9).

The mass-shell condition $p_t^2 =p_x^2$ provides
\begin{equation}
J_0^2 - J_1^2 - J_2^2 = 0,
\end{equation}
which defines two cones.

The physical conditions $\dot{t}>0$ and $\lambda >0$
give $p_t<0$. Thus, the upper-cone ($J_0>0$) without the vertex 
$J_0=0=J_1=J_2$ is the physical phase-space of our system.
By (4.5)  we get
\begin{equation}
J_0 \cos mx = J_1\cos mt - J_2 \sin mt .
\end{equation}
Due to  (4.6), the trajectories of massless particle (4.7) are
zigzag lines with the slope $\dot x =\pm1$.
A particle with the velocity equal to one reaches the
`edge' of space  where it `changes' the direction.
Then, it reaches another `edge', again changes the direction, etc.
 This motion is periodic with the period $2\pi /m$. 

Since the physical phase-space is the upper-cone without the vertex, 
the system can be quantized as in the case of
${\cal C}_+$ system, presented in Section 2. 
The corresponding quantum theory is defined by the 
$D_1^+$  representation.

\setcounter{equation}{0}
\section{Discussion}

Any two  Lorentzian 2d manifolds with the same constant curvature $R_0$
are (locally) isometric. Therefore, the considered systems give
the exhaustive picture for the dynamics of massless particle in 
2d spacetime with constant curvature.

It is interesting to note that these mechanical systems 
are related to the model of 2d gravity with the dilaton field [11].
In the conformal gauge this 2d gravity model is described by the Liouville
and free fields, which have equal traceless energy-momentum
tensors. One can check that the
massless field, which has the same traceless energy-momentum tensor 
as the Liouville field (4.2), is singular and
the singularity lines coincide with the
trajectories of the massless particle given by (4.7).
A similar picture exists for the hyperboloid (2.1) described by the
Liouville field (2.5).
The Hamiltonian reduction to the physical variables 
eliminates all field degrees of freedom of the indicated model of 2d gravity
and the physical phase-space of the model is finite dimensional.
The character of the reduced system and its relation
to the considered dynamics of massless particle
depends on the global properties of spacetime.
Details concerning this relation will be discussed elsewhere.  

{\bf Acknowledgments   }
This work was supported by the grants from:
INTAS  (96-0482), RFBR (96-01-00344),
the Georgian Academy of Sciences and the
So{\l}tan Institute for Nuclear Studies.

We wish to thank Mikhail Plyushchay for correspondence.

\end{document}